\documentclass{elsart}

\usepackage{graphicx}
\usepackage{amsmath}

\newcommand{\ket}[1]{|#1\rangle}

\begin{document}

\begin{frontmatter}
\bibliographystyle{elsart-num}

\title{Local basis-dependent noise induced Bell-nonlocality sudden death \\
in tripartite systems}

\author[Jaeger]{Gregg Jaeger} and
\ead{jaeger@bu.edu, correspondence author}
\author[Ann]{Kevin Ann}
\ead{kevinann@bu.edu}

\address[Jaeger]{Quantum Imaging Lab, Department of Electrical and Computer Engineering,\\
and Division of Natural Sciences, Boston University, Boston, MA
02215}
\address[Ann]{Department of Physics, Boston University \\
590 Commonwealth Avenue, Boston, MA 02215}

\date{\today}

\begin{abstract}
We demonstrate that multipartite Bell-inequality violations can be
fully destroyed in finite time in three-qubit systems subject only
to the mechanism of local external asymptotic dephasing noise. This
broadens the study of local-noise-induced sudden death of nonlocal
behavior, extending it beyond the realm of bipartite systems, to
which it had previously been restricted.
\end{abstract}


\end{frontmatter}

\section{Introduction}

It has been increasingly recognized that the rates of loss of
joint-state coherence and of entanglement, two fundamental
characteristics of quantum states, may differ within a composite
system subject to local external noise
\cite{YE03,AJ07a,AJ07b,DH04,YE04,YE06a,YE06b,YE07,AJ07c,AJ07d}.
Moreover, it has been shown that for some classes of states
entanglement sudden death (ESD) \cite{YE07}, the disentanglement of
bipartite systems in \emph{finite time} subject only to the
mechanism of basis-dependent local phase noise, occurs. Thus,
qualitative as well as quantitative differences in coherence and
nonlocality have been demonstrated
\cite{DH04,YE04,YE06a,YE06b,YE07,AJ07c,AJ07d}. Here, this
investigation is further advanced.

Previously, such differences of noise-induced behavior have been
carefully explored only in bipartite systems. The study of ESD under
local dephasing noise has not been demonstrated in any multipartite
system of more than two components because properly quantifying
multipartite entanglement, particularly for mixed states which it
involves by necessity, is problematic for systems of more than two
qudits \cite{Betal,BPRST}. Nonetheless, as we demonstrate here, one
may still demonstrate the existence of local-noise induced death of
nonlocal behavior with tools currently at hand. In particular, one
can find classes of states in which generalized Bell-nonlocality in
multipartite systems can go to zero in finite time while state
coherence continues to be maintained for all finite times, an effect
which we term {\it Bell-nonlocality sudden death (BNSD)}.  Here,
BNSD is demonstrated in the tripartite context: the destruction of
nonlocality as measured by the extent of violation of tripartite
Bell inequalities in finite time under basis-dependent multi-local
asymptotic dephasing noise is demonstrated in a class of initially
Bell nonlocal pure states of three-qubit systems, namely, the W
class of states.

The extension of the consideration of the sudden death of nonlocal
properties due to local dephasing noise beyond the bipartite case to
the tripartite case is important because tripartite systems can
exhibit fundamental characteristics impossible in bipartite systems,
even at the three-qubit level, for example, see \cite{GHZ,GHZ2}. The
exhibition of BNSD illuminates the quantum--classical transition,
quantum measurement, and quantum information processing where
joint-state coherence and nonlocality are typically considered
crucial. Our demonstration of BNSD shows that quantum information
processing may be even more challenging to carry out in a noisy
environment than previously thought.

\section{Model: initial state and noise model}
There are two distinct classes of entangled pure states for
three-qubit systems, the GHZ class and the W class, each represented
by a characteristic state related to all others of their class by
stochastic local operations and classical communication
\cite{DVC00}. Here, we restrict our attention to systems initially
prepared in pure states of the W class in order to show that
tripartite Bell inequality violations in such systems can be
eliminated by the mechanical influence of local asymptotic dephasing
noise alone.

The general pure W state is given by $\ket{\rm W_g} =
\bar{a}_{1}\ket{001} + \bar{a}_{2}\ket{010} + \bar{a}_{4}\ket{100},$
where $\bar{a}_i\in\mathcal{C}$ and $\sum_{i}|\bar{a}_i|^2=1$, and has
the associated density matrix
\begin{eqnarray}
\rho_{\rm Wg}=
\left(
\begin{array}{cccccccc}
 \ 0 \ &\ 0 \ &\ 0 \ &\  0\ \ &\ 0 \ &\  0\ & \ 0 \ &\  0\ \\
 0 & |\bar{a}_{1}|^2 & \bar{a}_{1}\bar{a}_{2}^* & 0 & \bar{a}_{1}\bar{a}_{4}^* & 0 & 0 & 0 \\
 0 & \bar{a}_{2}\bar{a}_{1}^* & |\bar{a}_{2}|^2 & 0 & \bar{a}_{2}\bar{a}_{4}^* & 0 & 0 & 0 \\
 0 & 0 & 0 & 0 & 0 & 0 & 0 & 0 \\
 0 & \bar{a}_{4}\bar{a}_{1}^* & \bar{a}_{4}\bar{a}_{2}^* & 0 & |\bar{a}_{4}|^2 & 0 & 0 & 0 \\
 0 & 0 & 0 & 0 & 0 & 0 & 0 & 0 \\
 0 & 0 & 0 & 0 & 0 & 0 & 0 & 0 \\
 0 & 0 & 0 & 0 & 0 & 0 & 0 & 0
\end{array}
\right).
\end{eqnarray}

In order to demonstrate the existence of BNSD in multipartite
systems, we model pure dephasing noise acting locally on each of the
three qubits which are assumed to be isolated from each other. The
most general time-evolved open-system density matrix expressible in
the operator-sum decomposition is
\begin{equation}
\rho\left(t\right) = \mathcal{E}\left(\rho\left(0\right)\right) =
\sum_{\mu}D_{\mu}\left(t\right)\rho\left(0\right)
D_{\mu}^{\dagger}\left(t\right) \ ,
\end{equation}
where the $D_\mu(t)$, which satisfy a completeness condition
guaranteeing that the evolution be trace-preserving, represent the
influence of local statistical noise, and where the index $\mu$ runs
over the number of elements required for the decomposition
\cite{Kraus83}. For local and multi-local dephasing environments,
the $D_{\mu}(t)$ are of the form $G_{k}(t)F_{j}(t)E_{i}(t)$, so that
\begin{eqnarray}
\rho\left(t\right) &=& \mathcal{E}\left(\rho\left(0\right)\right) =
\sum_{i = 1}^{2}\sum_{j = 1}^{2}\sum_{k = 1}^{2}
G_{k}\left(t\right)F_{j}\left(t\right)E_{i}\left(t\right)
\rho\left(0\right)
E_{i}^{\dagger}\left(t\right)F_{j}^{\dagger}\left(t\right)G_{k}^{\dagger}\left(t\right)\
, \nonumber \label{krausSpecific}
\end{eqnarray}
where
\begin{eqnarray}
E_{1}(t) &=& {\rm diag}(1,\gamma_{\rm A}(t)) \otimes \mathbf{I} \otimes \mathbf{I} \ \ , \ \
E_{2}(t)  = {\rm diag}(0,\omega_{\rm A}(t)) \otimes \mathbf{I} \otimes \mathbf{I} \ ,  \\
F_{1}(t) &=& \mathbf{I} \otimes {\rm diag}(1,\gamma_{\rm B}(t)) \otimes \mathbf{I} \ \ , \ \
F_{2}(t)  =  \mathbf{I} \otimes {\rm diag}(0,\omega_{\rm B}(t)) \otimes \mathbf{I} \ , \\
G_{1}(t) &=& \mathbf{I} \otimes \mathbf{I} \otimes {\rm diag}(1,\gamma_{\rm C}(t)) \ \ , \ \
G_{2}(t)  =  \mathbf{I} \otimes \mathbf{I} \otimes {\rm diag}(0,\omega_{\rm C}(t)) \ , \
\end{eqnarray}
$\gamma_{\rm A}\left(t\right) = \gamma_{\rm B}\left(t\right) = \gamma_{\rm C}\left(t\right) =
 \gamma\left(t\right) = e^{-\Gamma t},
 \omega_{\rm A}\left(t\right) = \omega_{\rm B}\left(t\right) = \omega_{\rm C}\left(t\right) =
 \omega\left(t\right) = \sqrt{1-\gamma^{2}(t)} = \sqrt{1-e^{- 2 \Gamma t}}$.
The $E_{i}(t)$, $F_{j}(t)$, and $G_{k}(t)$ induce dephasing in the
state of each qubit alone, and individually satisfy the usual
completeness condition for the operator-sum decomposition of CPTP
maps \cite{Kraus83}.  $\Gamma$ quantifies the rate of local
asymptotic dephasing which we assume to be equal for all qubits. We
take the time-dependence of $\gamma(t)$'s to be implicit from here
on, particularly when explicitly displaying the forms of density
matrices.

\section{Bell-nonlocality Sudden Death}

In the multi-local noise environment described in the previous
section, when a three-qubit system is prepared at the initial time
$t=0$ in the generic pure W state $\rho(0)=|{\rm
W_g}\rangle\langle{\rm W_g}|$ , the time-evolved state $\rho(t)$,
that is, the solution of Eq. (2) is
\begin{eqnarray}
\rho\left(t\right)= \left(
\begin{array}{cccccccc}
 0 &\ 0\ &\ 0\ &\ 0\ &\ 0\ &\ \ 0\ \ &\ \ 0\ \ &\ \ 0\ \ \\
 0 & |\bar{a}_{1}|^2 & \bar{a}_{1}\bar{a}_{2}^* \gamma_{\rm B}\gamma_{\rm C} & 0 & \bar{a}_{1}\bar{a}_{4}^* \gamma_{\rm A}\gamma_{\rm C}& 0 & 0 & 0 \\
 0 & \bar{a}_{2}\bar{a}_{1}^* \gamma_{\rm B}\gamma_{\rm C} & |\bar{a}_{2}|^2 & 0 & \bar{a}_{2}\bar{a}_{4}^* \gamma_{\rm A}\gamma_{\rm B} & 0 & 0 & 0 \\
 0 & 0 & 0 & 0 & 0 & 0 & 0 & 0 \\
 0 & \bar{a}_{4}\bar{a}_{1}^* \gamma_{\rm A}\gamma_{\rm C} & \bar{a}_{4}\bar{a}_{2}^* \gamma_{\rm A}\gamma_{\rm B}& 0 & |\bar{a}_{4}|^2 & 0 & 0 & 0 \\
 0 & 0 & 0 & 0 & 0 & 0 & 0 & 0 \\
 0 & 0 & 0 & 0 & 0 & 0 & 0 & 0 \\
 0 & 0 & 0 & 0 & 0 & 0 & 0 & 0
\end{array}
\right).
\end{eqnarray}

\noindent The off-diagonal elements of this matrix undergo a simple
exponential decay, the three-qubit state fully decohering only in
the infinite-time limit, wherein the $\gamma$ factors approach zero.
As we now show, however, the tripartite Bell nonlocality of states
is entirely lost in a specific finite time-scale.

For $N$-qubit systems, there are two multipartite Bell inequalities
that have been used to detect the degree of nonlocality as measured
by the extent of their violation. Satisfaction of the first, that of
Zukowski and Brukner \cite{ZB02}, is both a necessary and a
sufficient condition for a quantum system of $N$ qubits obey Bell
locality but is often difficult computationally to determine; it is
currently unclear whether a calculation valid for all classes of
states for $N>3$ can be carried out. The second inequality used for
this purpose is the Mermin-Ardehali-Belinksii-Klyshko (MABK)
inequality \cite{Mermin90,Ardehali92,BK93}, which may fail to detect
the existence of nonlocal correlation in some states for cases where
$N>3$ but always detects nonlocal correlation when $N=3$, has the
advantage of being readily computable. Because these two
inequalities are equivalent in the case $N=3$, we have a definitive
criterion for Bell locality valid for all three-qubit states, the
extent of the violation of which allows us to show its death.

A quantum state $\rho$ violates the MABK inequality if
\begin{equation}
\left|\left\langle \mathcal{B}_{N} \right\rangle_{\rho}\right| > 1.
\end{equation}
In the simplest case of two qubits, $N=2$, the operator
$\mathcal{B}_N$ on the left-hand side of the MABK inequality is just
the Clauser-Horne-Shimony-Holt (CHSH) operator, namely,
\begin{equation}
\mathcal{B}_{2} = \frac{1}{2}\left[M_{\rm A}M_{\rm B} + M_{\rm
A}M_{\rm B}' + M_{\rm A}'M_{\rm B} - M_{\rm A}'M_{\rm B}'\right].
\end{equation}

In the case of current interest, that of three qubits (here labeled
K=A,B, or C), the specific form of the operator $\mathcal{B}_N$ is
\begin{equation}
\mathcal{B}_{3} = \frac{1}{2}\left[M_{\rm A}M_{\rm B}M_{\rm C}' +
M_{\rm A}M_{\rm B}'M_{\rm C} + M_{\rm A}'M_{\rm B}M_{\rm C} - M_{\rm
A}'M_{\rm B}'M_{\rm C}' \right],
\end{equation}
the primed and unprimed terms denoting the two different directions
in which the corresponding party measures; the measurement operators
$M_{\rm K}$ and $M_{\rm K}'$, denoting the operators corresponding
to measurements on the qubit K, where the second corresponds to a
measurement performed along a direction differing by $\theta$
relative that performed on the first qubit, are
\begin{eqnarray}
\left(
\begin{array}{c}
 M_{\rm K} \\
 M_{\rm K}'
\end{array}
\right) =R(\theta_{\rm K})\left(
\begin{array}{c}
 M_{\rm A} \\
 M_{\rm A}' \end{array}
\right),
\end{eqnarray}

where
$R\left(\theta_{\rm K}\right) = \left(
\begin{array}{cc}
 \cos \theta_{\rm K} & -\sin \theta_{\rm K}  \\
 \sin \theta_{\rm K} & \ \ \cos \theta_{\rm K}
\end{array}
\right) \ .$

In the case of three qubits, the relative angles of measurement are
$\theta_{\rm B}=\pi / 6$ and $\theta_{\rm C}= \pi / 3$. Thus, the
corresponding measurement operators for qubits A, B, and C are
written in terms of the usual Pauli operators $\sigma_{z}$ and
$\sigma_{x}$, as
\begin{eqnarray}
M_{\rm A} &=& \sigma_{z} \otimes \mathbf{I} \otimes \mathbf{I} \ , \\
M_{\rm A}' &=& \sigma_{x} \otimes \mathbf{I} \otimes \mathbf{I} \ , \\
M_{\rm B}  &=& \mathbf{I} \otimes
\left[\cos\left(\frac{\pi}{6}\right)\sigma_{z}-\sin\left(\frac{\pi}{6}\right)\sigma_{x}\right]
\otimes \mathbf{I} \ , \\
M_{\rm B}' &=& \mathbf{I} \otimes
\left[\sin\left(\frac{\pi}{6}\right)\sigma_{z}+\cos\left(\frac{\pi}{6}\right)\sigma_{x}\right]
\otimes \mathbf{I} \ , \\
M_{\rm C}  &=& \mathbf{I} \otimes \mathbf{I} \otimes
\left[\cos\left(\frac{\pi}{3}\right)\sigma_{z}-\sin\left(\frac{\pi}{3}\right)\sigma_{x}\right] \ , \\
M_{\rm C}' &=& \mathbf{I} \otimes \mathbf{I} \otimes
\left[\sin\left(\frac{\pi}{3}\right)\sigma_{z}+\cos\left(\frac{\pi}{3}\right)\sigma_{x}\right]
\ .
\end{eqnarray}

The expectation value of the $\mathcal{B}_3$ operator for the state
under the influence of multi-local noise on three qubits initially
prepared in the generic pure W state is
\begin{eqnarray}
\left\langle \mathcal{B}_{3} \right\rangle_{\rho(t)} &=&
{\rm tr} \left[\mathcal{B}_{3}\left(t\right) \rho(t)\right] \nonumber \\
&=& {\rm tr} \left[ \frac{1}{2}\big(M_{\rm A}M_{\rm B}M_{\rm C}' +
M_{\rm A}M_{\rm B}'M_{\rm C} + M_{\rm A}'M_{\rm B}M_{\rm C} - M_{\rm
A}'M_{\rm B}'M_{\rm C}' \big)
\rho(t) \right] \nonumber \\
&=& \frac{1}{2}{\rm tr}\left[M_{\rm A}M_{\rm B}M_{\rm
C}'\rho(t)\right] + \frac{1}{2}{\rm tr}\left[M_{\rm A}M_{\rm
B}'M_{\rm C}\rho(t)\right] +
    \frac{1}{2}{\rm tr}\left[M_{\rm A}'M_{\rm B}M_{\rm C}\rho(t)\right]\nonumber \\
    &&\ \ \ \ \ \ - \frac{1}{2}{\rm tr}\left[M_{\rm A}'M_{\rm B}'M_{\rm C}'\rho(t)\right] \nonumber\ . \\
\end{eqnarray}
Therefore,
one finds
\begin{eqnarray}
\frac{1}{2}{\rm tr}\left[M_{\rm A}M_{\rm B}M_{\rm C}'\rho(t)\right]
&=& - \frac{1}{8} \left[3 + \left( \bar{a}_{1}\bar{a}_{2}^{\ast} +
\bar{a}_{2}\bar{a}_{1}^{\ast} \right)
\gamma_{\rm B}\gamma_{\rm C} \right]\ , \\
\frac{1}{2}{\rm tr}\left[M_{\rm A}M_{\rm B}'M_{\rm C}\rho(t)\right]
&=&- \frac{1}{8} \left[1 + 3 \left( \bar{a}_{1}\bar{a}_{2}^{\ast} +
\bar{a}_{2}\bar{a}_{1}^{\ast} \right)
\gamma_{\rm B}\gamma_{\rm C} \right]\ , \\
\frac{1}{2}{\rm tr}\left[M_{\rm A}'M_{\rm B}M_{\rm C}\rho(t)\right]
&=&- \frac{1}{8} \gamma_{\rm A}\left[\left(
\bar{a}_{2}\bar{a}_{4}^{\ast} + \bar{a}_{4}\bar{a}_{2}^{\ast}
\right)\gamma_{\rm B}
+ 3 \left( \bar{a}_{1}\bar{a}_{4}^{\ast} + \bar{a}_{4}\bar{a}_{1}^{\ast} \right)\gamma_{C} \right]\ , \\
\frac{1}{2}{\rm tr}\left[M_{\rm A}'M_{\rm B}'M_{\rm
C}'\rho(t)\right]&=& \frac{1}{8} \gamma_{\rm A}\left[3 \left(
\bar{a}_{2}\bar{a}_{4}^{\ast} + \bar{a}_{4}\bar{a}_{2}^{\ast}
\right)\gamma_{B} + \left( \bar{a}_{1}\bar{a}_{4}^{\ast} +
\bar{a}_{4}\bar{a}_{1}^{\ast} \right)\gamma_{\rm C} \right]\ .
\end{eqnarray}
We then have
\begin{eqnarray}
|\langle \mathcal{B}_{3} \rangle_{\rho(t)}| = \frac{1}{2}\big|1 + (
\bar{a}_{1}\bar{a}_{2}^{\ast} +
\bar{a}_{2}\bar{a}_{1}^{\ast})\gamma_{B}\gamma_{C} + (
&\bar{a}_{2}&\bar{a}_{4}^{\ast} +
\bar{a}_{4}\bar{a}_{2}^{\ast})\gamma_{A}\gamma_{B}\nonumber\\
+(&\bar{a}_{1}&\bar{a}_{4}^{\ast} +
\bar{a}_{4}\bar{a}_{1}^{\ast})\gamma_{A}\gamma_{C}\big|\ .
\end{eqnarray}

Thus, tripartite Bell nonlocality is nonexistent by the time-scale
\begin{equation}
\tau_{\rm BNSD} = \ln(\bar{a}_{1}\bar{a}_{2}^{\ast} +
\bar{a}_{2}\bar{a}_{1}^{\ast} + \bar{a}_{2}\bar{a}_{4}^{\ast} +
\bar{a}_{4}\bar{a}_{2}^{\ast} + \bar{a}_{1}\bar{a}_{4}^{\ast} +
\bar{a}_{4}\bar{a}_{1}^{\ast}) / 2 \Gamma
\end{equation}
specified by the time for which $\left|\left\langle
\mathcal{B}_{3}\right\rangle_{\rho\left(t\right)}\right|$ reaches
$1$ from above. This suffices to demonstrate tripartite
Bell-nonlocality sudden death of initially Bell-nonlocal states
under local dephasing noise: by Eq. (23), whenever $1/2 < a_1 a_2 +
a_2 a_4 + a_1 a_4$ we have $\left|\left\langle \mathcal{B}_{3}
\right\rangle_{\rm Wg}\right|$ $> 1$, a violation of the MABK
inequality, and in the limit $t \rightarrow \infty$
$|\left\langle\mathcal{B}_{3}\right\rangle_{\rho(t)}|\rightarrow 1 /
2$, no longer violating the inequality. In particular, consider the
particular case in which all amplitudes are equal and real, that is,
the standard W state. Then we find that both $a_1 a_2 + a_2 a_4 +
a_1 a_4 = 1$ and
\begin{equation} \tau_{\rm
BNSD} = \frac{\ln (2)}{2 \Gamma} \ .
\end{equation}
\section{Conclusions}

We have demonstrated the destruction of Bell-nonlocal behavior under
basis-dependent local asymptotic dephasing noise, as measured by the
extent of violation of a tripartite Bell inequality, in finite time
while state coherence remains for all finite times in a class of
initial states of three-qubit systems. This illuminates the
quantum--classical transition, and quantum information processing in
particular, because it shows in the multipartite context that
nonlocal behavior can be lost simply due to the influence of local
asymptotic dephasing noise, that in which practical quantum
computing and quantum measurement will typically take place, despite
the persistence of simple coherence for all finite times,
strengthening the view that environmental noise makes quantum
information processing a particularly challenging task.

\end{document}